# Metal hardening in atomistic detail§

Luis A. Zepeda-Ruiz[1], Alexander Stukowski[2], Tomas Oppelstrup[1], Nicolas Bertin[1], Nathan R. Barton[1], Rodrigo Freitas[3,4], Vasily V. Bulatov[1]*

[1]Lawrence Livermore National Laboratory.
[2]Technische Universität Darmstadt.
[3]University of California Berkeley.
[4]Stanford University.

*Correspondence to: bulatov1@llnl.gov

§*Dedicated to the memory of Ali Argon*.

Abstract

Through millennia humans exploited the natural property of metals to get stronger or hardened when mechanically deformed. Ultimately rooted in the motion of dislocations, mechanisms of metal hardening remained in the crosshairs of physical metallurgists for over a century. Here, we performed atomistic simulations at the limits of supercomputing, which are sufficiently large to be statistically representative of macroscopic crystal plasticity yet fully resolved to examine the origins of metal hardening at its most fundamental level of atomic motion. We demonstrate that the notorious staged (inflection) hardening of metals is a direct consequence of crystal rotation under uniaxial straining. At variance with widely divergent and contradictory views in the literature, we observe that basic mechanisms of dislocation behavior are the same across all stages of metal hardening.

Root causes of metal hardening remained unknown until 86 years ago when dislocations – curvilinear crystal defects made of lattice disorder – were proposed to be responsible for crystal plasticity. Yet, even though direct causal connection between dislocations and crystal plasticity is now firmly established, no quantitative theory exists to predict metal hardening directly from the underlying behavior of lattice dislocations. This has not been for lack of trying as numerous theories and models sprang to life that explained metal hardening from dislocation mechanisms, often based on differing if not opposing viewpoints[1]. But as physicist Alan Cottrell opined, work (strain) hardening is perhaps the most difficult remaining problem in classical physics (worse than turbulence) and is likely to be solved last[2]. The essential difficulty in putting to rest still-ongoing debates on mechanisms of strain hardening has been our persistent inability to observe what dislocations do *in situ* – during straining – in the material bulk.

While physicists grapple to understand and quantify strain hardening, material models employed to optimize metal processing (e.g. forging, rolling or extrusion) remain phenomenological and based on empirical observations largely predating the very concept of crystal dislocations. Here we rely on a super-computer to clarify what causes metal hardening. Instead of trying to derive hardening from the underlying mechanisms of dislocation behavior, which has been the aspiration of dislocation theory for decades, we perform ultra-scale computer simulations at a still more basic level – the motion of atoms that the crystal is made of. Viewed here as unbiased computational experiments, in our simulations we observe, rather than prescribe, exactly how the motion of individual atoms translates into the motion of dislocations that then conspire to produce metal hardening. Here we focus our attention on clarifying the origins of so-called "three-stage" hardening as perhaps most argued-about aspect of metal plasticity.

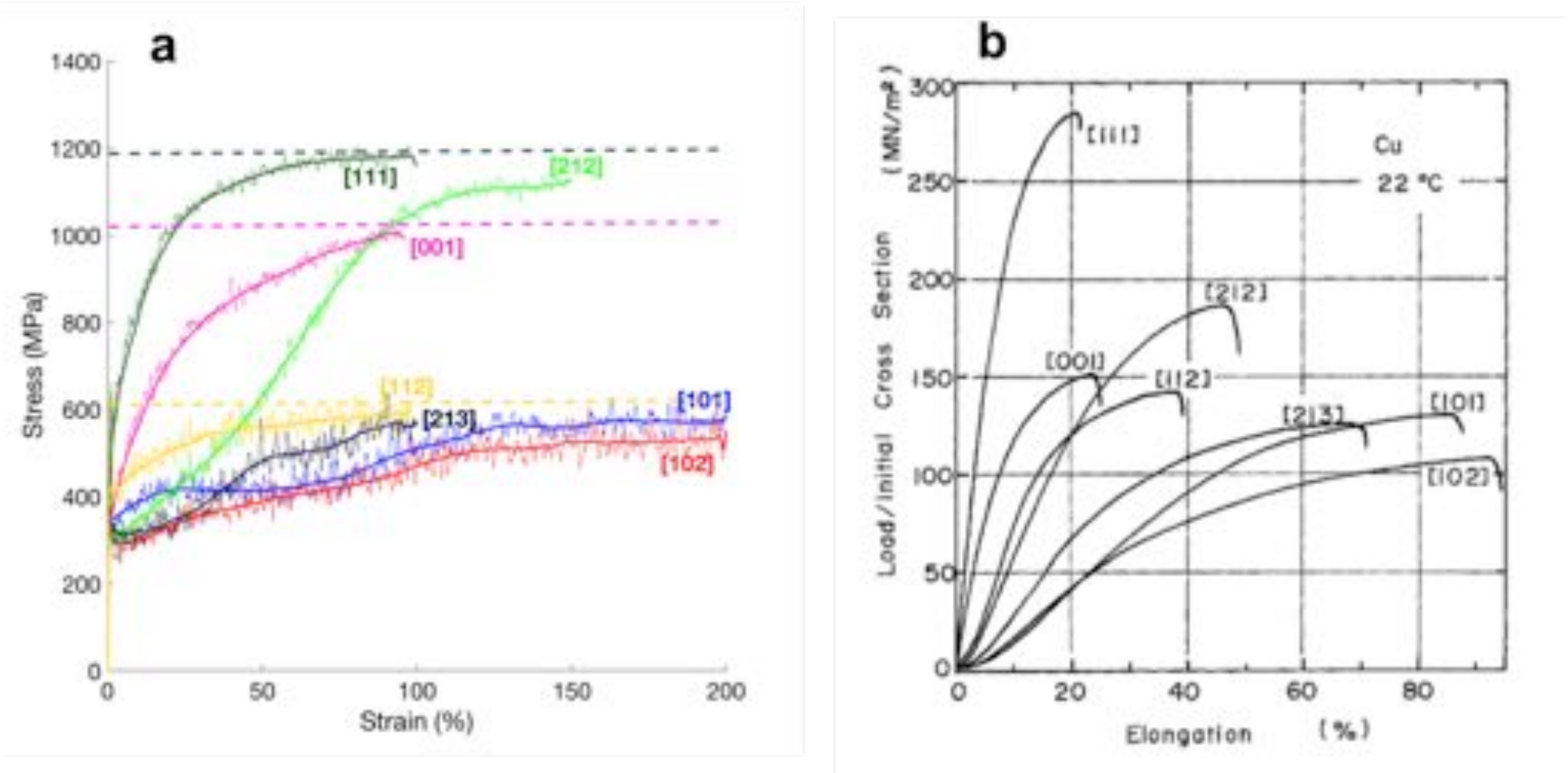

**Fig. 1**: **Stress-strain response of an aluminum single crystal subjected to tensile straining along seven different initial orientations of the straining axis. a**, Stress-strain response extracted from MD simulations. Each line is labeled with the Miller index[29] of the crystal's initial axis orientation. The thin lines are raw stress-strain data, the thick lines are the same data smoothed using a moving average filter. The horizontal dashed lines are flow stress levels attained asymptotically as crystals approach their stable end orientations. **b**, Corresponding experimental stress-strain curves obtained in tensile straining tests of single crystal copper[3].

Figure 1a shows stress-strain response observed in our molecular dynamics (MD) simulations of seven aluminum single crystals subjected to uniaxial tension at ambient temperature and pressure (see Methods for simulation details). Shown in Fig. 1b are results of uniaxial tension experiments on the same seven crystals reported by Takeuchi in his classic 1975 paper[3]. Leaving aside for now quantitative differences in the magnitude of the flow stress, the simulated and experimental stress-strain curves are in remarkable qualitative agreement. Indeed, relative ordering of the flow stress among seven simulated crystals is the same as in the Takeuchi's experiment both in the early as well as in the late stages of straining. Still more striking is the agreement in the shapes of the seven response curves: whereas three crystals labeled [001], [111] and [112] exhibit simple "parabolic" stress rise (hardening), the other four curves show more or less distinct inflection, or "three-stage" hardening. The difference among the seven crystals was in the orientations of their crystal lattices (with respect to the straining axis), selected here after Takeuchi so as to sample important lattice orientations of a cubic crystal. Essential for this work is that in his paper Takeuchi also reports if and how each of the seven crystals rotates in response to tensile straining[3]. Even though Takeuchi's results are for copper, qualitatively similar response has been widely reported in aluminum studied here and in most other face-centered cubic (FCC) metals[4-10]. Further justification for our choice of aluminum as a representative FCC metal is given in Supplementary Discussion 4 and Supplementary Figure 4.

The striking similarities between the two sets of stress-strain curves shown in Figs. 1a and 1b cannot be accidental despite an astonishing 10 orders of magnitude difference in straining rates: $5 \cdot 10^7$/s in our simulations and $3 \cdot 10^{-3}$/s in the experiment[3]. This agreement suggests that: (a) fundamental physical mechanisms of dislocation plasticity remain invariant across the entire range of straining rates from low or "quasi-static" rates of laboratory experiments ($10^{-5}$ – $10^1$/s) to high or "dynamic" rates of our simulations ($10^5$ – $10^8$/s); (b) viewed as computational experiments, our high-rate MD simulations are just as quasi-static as the real low-rate laboratory experiments. Seemingly contradictory, the latter proposition derives from the following insights.

Often applied to straining experiments at rates exceeding $10^1$/s, the moniker "dynamic", essentially means that straining conditions – straining rate, temperature, pressure, etc. – cannot be maintained stationary at such rates in a real laboratory test primarily due to a finite rate of heat exchange and to inevitably imperfect control over straining apparatuses. This is in contrast to low-rate "quasi-static" experiments in which straining conditions can and usually are maintained strictly stationary. Yet, despite very high rates of straining necessitated here by the notorious time limit of MD simulations, our "computational experiments" are essentially quasi-static as we employ special controls – "Maxwell's demons" of sorts – that react to and gently nudge the motion of individual atoms so as to maintain the overall temperature, pressure and straining rate stationary (see Methods). While the total duration of our simulated MD trajectories is in the range 20-40 ns, this is still far longer than characteristic time scales of dislocation motion that delineate truly dynamic from quasi-static response. For instance, it takes only a few ps for a dislocation to adjust its velocity to a change in local stress (dislocation inertia), or a fraction of one ns for the same dislocation to travel from one obstacle to another. Thus, even if flow stress, dislocation density and other rate-dependent measures of metal plasticity reported here are understandably different, we maintain that, with respect to the underlying dynamics of dislocation behavior, our high-rate MD simulations are just as quasi-static as the low-rate laboratory experiments. Thus our observations presented below should be representative of quasi-static metal plasticity across the entire range of accessible straining rates.

Our key observation is that staged hardening is a direct manifestation of crystal rotation. Listed in Table 1 in order of decreasing initial slip symmetry, characteristic variations in the shapes of stress-strain curves are directly attributable to the occurrence (or not) of crystal rotation during straining. Three-stage (inflection) hardening is observed in the five crystals that rotate under straining (see also Supplementary Discussion 3), whereas parabolic hardening without an inflection is observed in the curves of the three crystals that do not rotate.

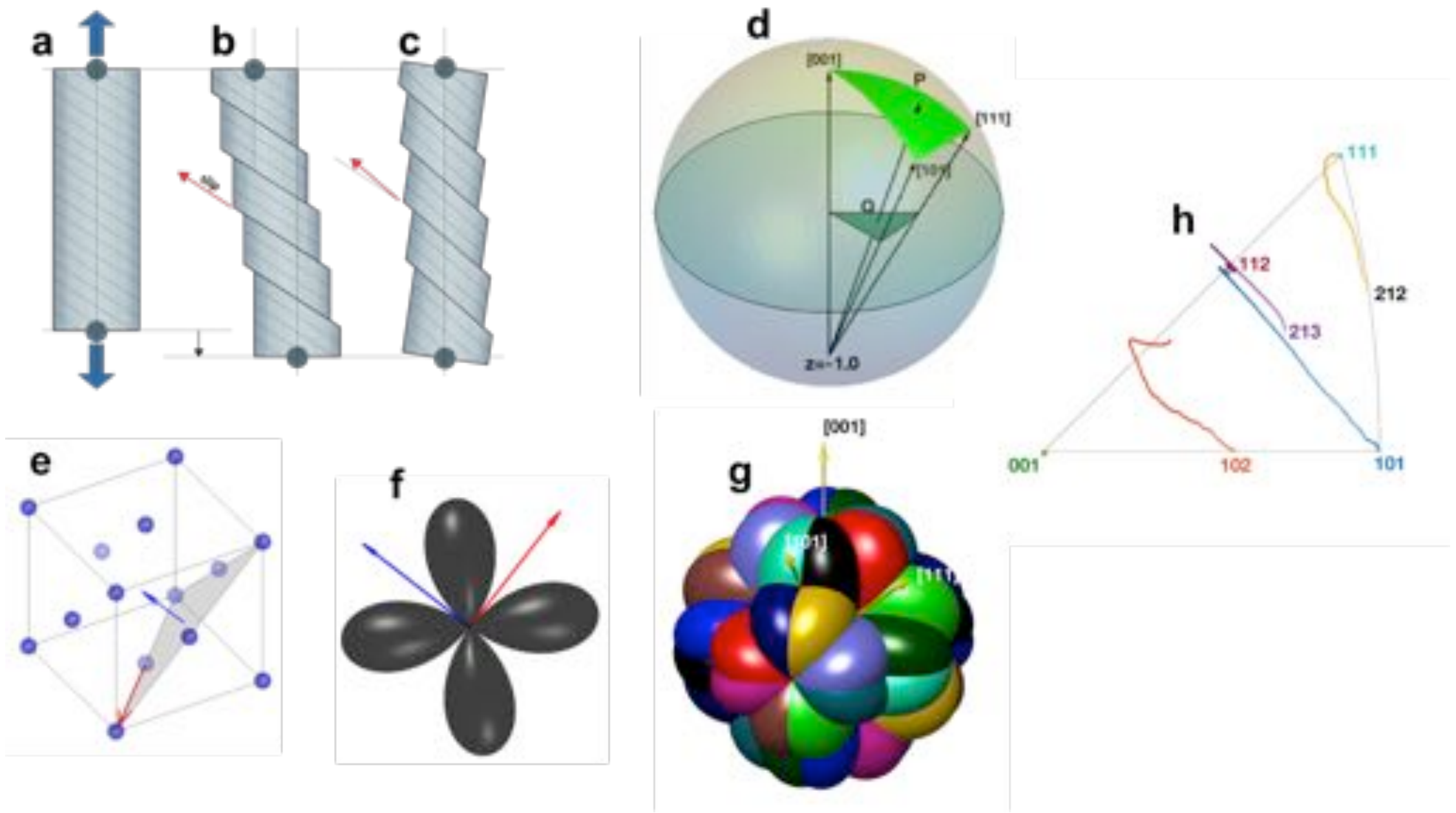

**Fig. 2**: **Slip crystallography of face-centered cubic single crystals. a**, **b**, **c**, Schematic illustration of crystal rotation due to slip on a single slip system under uniaxial tension. **d**, In the stereographic projection, point P representing an axis orientation on the unit sphere is projected onto point Q. A spherical triangle of axis orientations (bright green) is mapped onto a stereographic triangle on the equatorial plane (dark green). **e**, Slip direction (red arrow) and slip plane (gray) of one of the 12 slip systems in the face-centered cubic lattice. **f**, A 3D polar plot of the Schmid factor as a function of axis orientation with respect to the slip direction (red arrow) and the slip plane normal (blue arrow). **g**, The outer envelope of Schmid factors of all 12 slip systems in an FCC crystal under uniaxial straining. Plotted in 12 distinct colors, each of the 48 patches represents the amplitude of the Schmid factor on a slip system most favored for slip for a given axis orientation. The grooves coincide with the edges of the standard triangle

where two slip systems are equally favored. At the intersection of the grooves lie three types of cusp orientations in which four, six or eight slip systems are equally favored, corresponding to high-symmetry axis orientations. **h**, Rotation trajectories of seven crystals shown in Fig. 1a. Placed next to the starting point of each trajectory are the Miller indices of the initial axis orientation.

It turns out to be possible to explain why and, to a lesser extent, how crystals rotate under uniaxial straining without ever invoking lattice dislocations. About a century ago, well before lattice dislocations were first hypothesized[11-13] and later observed[14], it was discovered that single crystals, i.e. crystals with the same lattice orientation across the entire specimen volume, respond to straining by slipping in specific lattice planes along specific lattice directions[15-17]. Figure 2a shows that each such pair of a slip plane and a slip direction constitutes a slip system. In most crystals, several slip systems exist that can potentially contribute to macroscopic slip. Based on empirical observations, the celebrated Schmid law states that, when multiple slip systems exist, slip occurs in systems that are most favorably inclined with respect to the straining axis; this preference for slip is commonly expressed by the geometric Schmid factor[18]. Schmid also rationalized why some single crystal specimens rotate under uniaxial straining[19].

As shown in Fig. 2b, when orientation of the straining axis is such that one of the crystal's systems is most favorably inclined for slip, slipping along this primary slip vector should skew the specimen's shape. However, the specimen is constrained to stay co-axial with the grips of a stiff straining machine and compensates for its inability to skew by rotation, Fig. 2c. Based on such purely geometrical considerations, Schmid predicted that under tension a crystal should rotate so as to align its dominant slip direction with the straining axis[20]. Conversely, in a frame tied to the crystal lattice the straining axis rotates towards the dominant slip direction.

Takeuchi[3] selected his seven crystal orientations so as to sample all possible symmetries with respect to the straining axis of the face-centered cubic (FCC) lattice. To follow the logic, it is convenient to use the stereographic projection (Fig. 2d) and to map orientations of the straining axis onto the so-called standard triangle that constitutes just 1/48 of the unit sphere of axis orientations. Owing to the high symmetry of the cubic lattice, axis orientations within just one such triangle represent all other possible axis orientations. The standard triangle is particularly convenient for understanding slip crystallography in FCC crystals, because for every axis orientation in the interior of each triangle exactly one of the 12 slip systems – the primary system - is most favorably inclined for slip under uniaxial extension (Figs. 2e, 2f, 2g). Axis orientations on triangle edges are shared by two adjacent triangles and have at least two equally favored slip systems seeing the highest Schmid factor. Triangle corners [101], [111] and [001] lie at the intersections of two, three and four triangle edges and straining along these corner axes equally favors slip in four, six and eight slip systems, respectively. In Supplementary Discussion 1 we augment Schmid's geometric arguments with an *a priori* slip stability and reorientation flow analyses to predict which of the crystals studied here should rotate and which should not and, for each crystal predicted to rotate, where it should rotate to.

All but one crystal orientation behave precisely as predicted (Fig. 2h). Three crystals with axis orientations [001], [111] and [112] do not rotate at all. Crystals [213], [102], and [101] all eventually rotate to [112]. However, unlike crystal [213] that has no slip symmetry to begin with, crystals [102] and [101] break their initial 2-fold and 4-fold slip symmetries before they begin to rotate. Also predicted to break its 2-fold slip symmetry and then rotate to [112], crystal [212] instead retains its symmetry and rotates towards [111] along the [101]-[111] triangle edge. Even though we are yet to understand why this crystal rotates not as predicted, its rotation trajectory too agrees with Takeuchi's and other experiments[5]. This lends additional support to our observation that, despite their vastly different straining rates, our simulations and laboratory experiments probe the same physics of crystal plasticity. Since MD simulations allow unfettered access to every detail of atomic motion *in silico*, it should be possible to eventually establish what causes this and other still unexplained behaviors in metal plasticity.

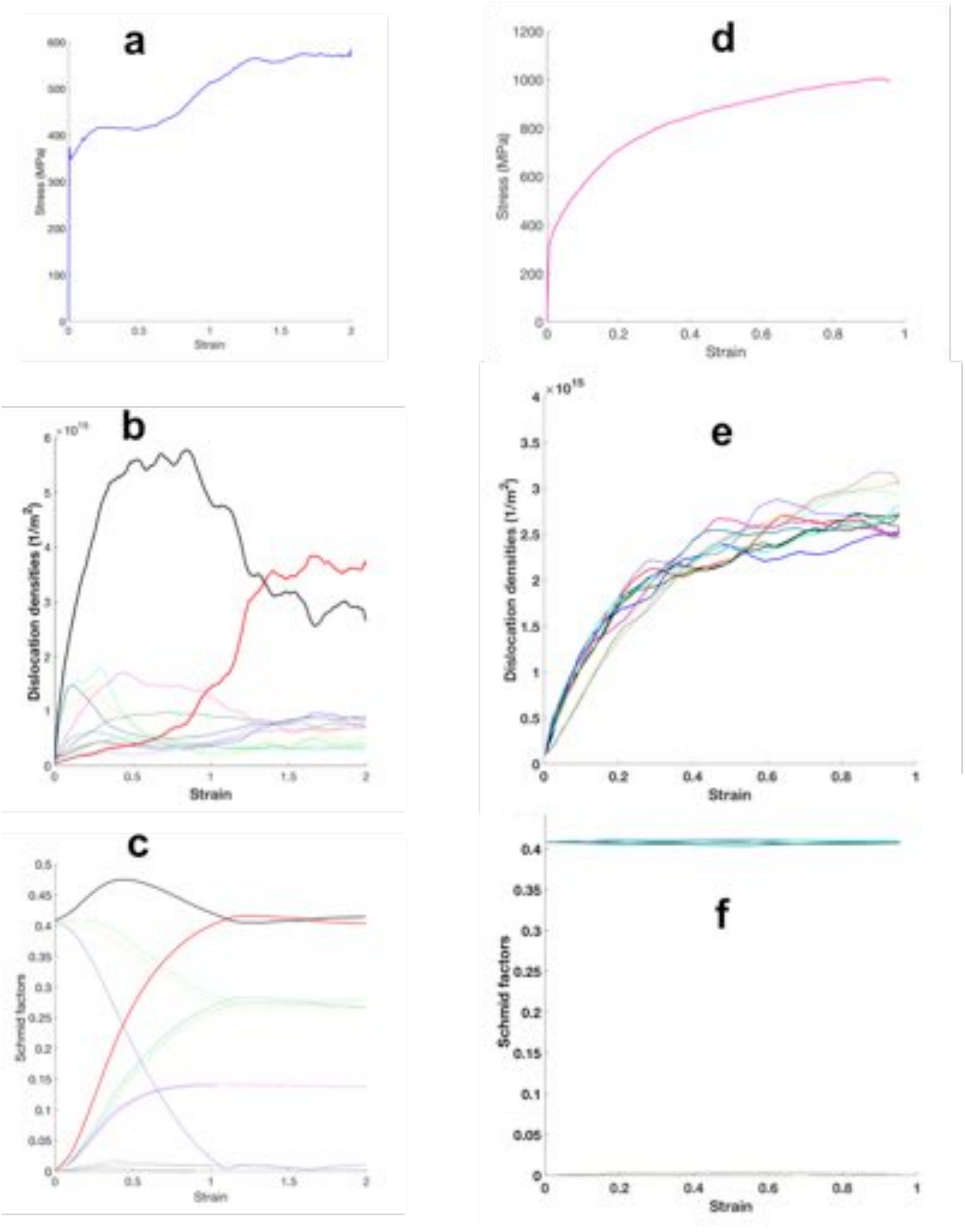

Fig. 3: **Stress-strain response, dislocations densities and Schmid factors in 12 slip systems of aluminum.** **a**, Stress versus strain, **b**, dislocation densities versus strain and **c**, Schmid factors versus strain computed under tensile straining of a crystal with straining axis initially aligned with [101]. Thick lines are densities of the primary and the conjugate slip systems. All other slip systems are shown as thin lines. **d**, **e**, **f**, The same curves for a crystal strained along its [001] axis. Thick lines are densities of eight primary systems and four inactive slip systems are shown in thin lines. Line colors in b, c, e and f are coordinated with each other and with the colors of corresponding spherical triangles in Fig. 2g.

To understand *why* and *how* crystal rotation causes staged hardening here we take advantage of our unique ability to see what dislocations do *in silico*, during straining in an MD simulation. Unlike mesoscale method of Discrete Dislocation Dynamics (DDD) that prescribes how dislocations respond to mechanical stress caused by straining[21,22], our approach is much more basic – we only specify how aluminum atoms interact with each other[23]. We then extract dislocations from atomistic configurations encountered every 0.1 ns along the simulated straining trajectories and partition the so-extracted dislocations over 12 slip systems of the FCC crystal (see Supplementary Discussion 2 for details). Figure 3 presents two out of seven straining simulations in which stress response, slip activity and Schmid factors of all 12 slip systems are coordinated along the straining trajectories (similar plots for five other simulated crystals are given in Supplementary Discussion 3).

The simulated straining response of an aluminum crystal initially oriented along the [101] axis is shown in Fig. 3a. Although all 12 slip systems were initially equally populated with dislocations (see Methods), the density of dislocations in the preferred four slip systems grows quickly from the start reflecting high slip activity in them, as shown in Fig. 3b. Owing to mutual forest resistance of intersecting dislocations in the four active systems the initial post-yield rate of hardening is high – when observed, such initially rapid hardening is sometimes referred to as stage-0 hardening. Soon the initial 4-fold symmetry breaks and one of the four systems (black line) continues to multiply dislocations while dislocation densities in three other initially favored systems (green, yellow and blue) subside. This causes the forest resistance and the hardening rate to drop entering what is called easy glide or stage I hardening. Just as predicted by Schmid, when one slip system dominates, the straining axis rotates towards its primary slip direction [011]. While the axis rotates across the triangle (Fig. 2h), the primary system becomes still more favored (note the maximum in its Schmid factor in Fig. 3c) so that the flow stress stays flat or even decreases slightly. On approaching the triangle edge, another “conjugate” system begins to see increasing Schmid factor (red line in Fig. 3c) and gradually activates. Dislocations in the conjugate system become more numerous increasingly blocking the motion of primary dislocations and causing the flow stress to rise more rapidly again due to increasing mutual forest resistance between the two active systems – this is hardening stage II. Schmid factors on the conjugate and primary systems become equal precisely when the axis crosses the triangle edge, but the axis trajectory slightly overshoots the edge and only then returns to it and settles at [112]. Observed in experiments, axis overshoot and subsequent axis oscillations are manifestations of “latent hardening” caused by delayed dislocation multiplication first in the conjugate system, then in the primary system, etc. As the axis settles in its end [112] orientation, both the flow stress and the fractional dislocation densities gradually approach their asymptotic values, with the associated reduction in hardening rate marking a transition to stage III. As was argued before[1] and supported by our recent results[24], asymptotic saturation of the flow stress and dislocation density can be achieved when dislocation multiplication becomes balanced by dislocation annihilation. In real experiments such asymptotic saturation is invariably interrupted by a macroscopic catastrophe such as necking and fracture or barreling so that the impending flow stress saturation is hinted at but not fully resolved in the experimental stress-strain curves (Fig. 1b)[3]. Although previously reported in the literature, most of these dislocation activity and evolution behaviors in staged hardening were inferred from circumstantial evidence, such as surface slip trace analysis and *postmortem* electron microscopy. Having access to all of the details of simulated atomistic trajectories, we present three videos illustrating how much more insight can be gained by observing what dislocations do *in silico* in material bulk (Supplementary Movies 1-3).

Staged hardening observed in MD simulations of other crystal orientations is just as tightly connected to crystal rotation, however detailed evolution of slip systems observed in MD simulations varies from case to case (see Supplementary Discussion 3). Three tested crystals with orientations [001], [111] and [112] do not rotate, retain their initial symmetries, and show no staged hardening (Fig. 1a). Considering [001] straining as a representative example (Fig. 3d), evolution of dislocation populations in 12 slip systems observed in these three simulations is rather uneventful: all initially favored systems remain equally favored (Fig. 3f) and active through straining as is evidenced by a steady rise and eventual saturation of dislocation densities in all active slip systems (Supplementary Movie 1). A surprising result however is that dislocations in systems with zero Schmid factors multiply nearly as much or even more than the active systems. In particular, under [001] straining, asymptotic density produced in each of the four "inactive" systems exceeds asymptotic densities in each of the eight active systems (Fig. 3e). It was recently reported[25], generation of dislocations in nominally inactive systems via junction reactions is a potentially important contribution to latent hardening broadly defined as any effect of active systems on the behavior of inactive systems[26]. We expect such latent multiplication of "dislocation dark matter" to be consequential for metal plasticity.

In addition to eight simulations reported in Table 1 (seven of which are shown in Fig. 1a), we performed smaller (~ 40 million atoms) simulations of six additional initial axis orientations. Out of 14 crystal orientations, all but three crystals are observed to be unstable and, when strained to a sufficiently large strain, asymptotically rotate towards one of the three stable "attractor" orientations. Thus, all possible crystal orientations naturally partition themselves into non-overlapping conical patches in orientation space, each cone centered on one of the three attractor orientations. On approaching their respective end orientations, rotating crystals raise their symmetry to double-slip (all the ones rotating to [112]), to sextuple-slip (approaching [111]) or to octuple-slip (approaching [001]). Flow stress attained in each of the three asymptotic attractor orientations is maximum over all orientations from the same conical patch (three dashed lines in Fig. 1a): within their local cones of attraction the three stable orientations are both the strongest and most symmetric. Our analysis of axis stability (Supplementary Discussion 1) further predicts that the [001] orientation should be only marginally stable under tension. Fittingly, we are yet to find a single initial axis orientation to rotate towards [001]: the cone of orientations attracted to [001] appears to be narrow. The ability to simulate lattice reorientation at very large strains and to identify asymptotic attractor orientations in MD simulations is important for improving accuracy of engineering models used to predict evolution of polycrystalline texture in rolling, forging, extrusion, ECAP, and other metal-forming operations.

Evidence is abundant in the literature that staged hardening in single crystals is a general phenomenon not limited to aluminum, FCC metals or cubic crystals and is observed in metals, semiconductors and ionic crystals alike. Our simulations clarify that three-stage hardening is not an intrinsic material property, but a kinematic consequence of the co-axiality constraint imposed on specimens in the standard uniaxial tests. Thus, it makes little sense to seek explanations of staged hardening in dislocation mechanisms somehow changing from one hardening stage to the next. At the same time, our simulations bring to light several previously unknown and potentially important aspects of metal plasticity that require one to delve more deeply into the details of dislocation motion for an explanation. Having access to the entire MD trajectory and armed with our recently developed methods of *in silico* computational microscopy[28,28], we can now relate every wiggle in the stress-strain response to underlying events in the life of atoms and dislocations. If, as we posit, our high-rate MD simulations and low-rate experiments probe the same physics, simulations of the kind presented here offer unique means for inquiry into fundamentals of crystal plasticity.

# Methods

## Molecular Dynamics (MD) simulations

To overcome notoriously severe limits on length and time scales of MD simulations, we ran most simulations reported here on Mira, one of the world's most powerful massively parallel supercomputers at Argonne Laboratory Computational Facility (ALCF). Interaction among aluminum atoms was modeled with a well-known embedded atom method (EAM) interatomic potential[23] that accurately describes a number of relevant properties of aluminum including its relatively high stacking-fault energy, which is expected to result in a high rate of dislocation multiplication under straining. All simulations were performed using the opensource code LAMMPS[30] on periodic fragments of single-crystal aluminum seamlessly embedded into an infinite crystal under three dimensional periodic boundary conditions. To confine the crystal to remain co-axial with the straining axis, we aligned one of the periodic box edges with the positive x axis of the LAMMPS coordinate frame. Uniaxial straining was simulated by extending the so-aligned edge of the periodic box. According to a convention adopted in LAMMPS, if initially aligned with the LAMMPS x axis, a box edge vector remains aligned with x while two other box edge vectors can rotate. Thus, arbitrary triclinic strain can be imposed on the simulated crystal. Initially orthorhombic, the periodic simulation box was allowed to evolve during the simulations into a general triclinic shape so as to maintain all but the xx component of the internal stress tensor near zero, as is known to be the case under uniaxial tensile straining and is confirmed by our own FEM simulations of a dog bone specimen of single crystal aluminum (see Supplementary Discussion 4 and Supplementary Movie 4). The simulated crystals each contained about 300 million atoms and had initial aspect ratios close to 1:2:2. To simulate aluminum's intrinsic plastic response it would have been prudent to perform our simulations at a constant 'true' straining rate (constant rate of relative elongation). However, virtually all experiments, including the ones by Takeuchi[3], have been performed under constant 'engineering' rate (constant crosshead velocity). To facilitate comparisons with the experiments, tensile straining was induced in our simulations by extending the x edge of the periodic box at a constant engineering rate of $5 \cdot 10^7$/s to the total strain ranging between 100% and 200%. Dislocations were seeded in the simulation volumes by carving out rhomb-shaped prismatic loops of the vacancy type, six loops for each of the six FCC Burgers vectors of the ½<110> type. Rhomb edges were aligned along two <112> directions perpendicular to the loop's Burgers vector so as to place the resulting ½<110> edge dislocations on two proper {111} glide planes. Length of the loop edges was 56Å which made for an initial dislocation density of $1.5 \cdot 10^{14}/m^2$.

Temperature was maintained constant with the help of a Maxwell's demon - a Langevin thermostat as implemented in LAMMPS. The damping parameter (relaxation time) of the thermostat was chosen so as to maintain temperature constant at 300K by removing heat generated by dislocation motion while not distorting atomic trajectories too much. In the range of straining rates employed here a damping parameter of 10-20 ps appears to be appropriate. To maintain uniaxial stress conditions characteristic of crystals subjected to tensile straining we employed yet another Maxwell's demon – a Parrinello-Rahman barostat[31] as implemented in LAMMPS *fix nph*. Again, to be gentle on the atomic trajectories, relaxation time constant of the barostat was set to 10 ps. Everywhere in the text by 'stress' we mean the uniaxial tensile Cauchy stress $\sigma_{XX}$ that develops in the model crystals in response to straining along the x axis.

***In silico* computational microscopy**

MD simulations of the kind reported here generate more than 1 exabyte (= $10^{18}$ bytes) of transient atomic trajectory data in just 12 hours on the Mira supercomputer. To be useable, data streams so immense must be compressed and recast into a form a human can comprehend. Here we rely on our recently developed methods of *in silico* computational microscopy to identify and precisely characterize dislocations in crystals[28,29] and to reduce the amount of data by three orders of magnitude. Structural defects in the simulated crystals were revealed using the dislocation extraction algorithm (DXA), a method that traces and indexes the dislocation lines and builds a concise network representation of the dislocation microstructure. If desired dislocations can be extracted at every time step, however for Supplementary Movies 1-3 we chose to perform DXA extraction at regular time intervals 0.1 ns apart which was frequent enough to follow the motion and reactions of dislocations in sufficient detail for our analyses.

Supplementary Movies 1-3 illustrate how much more insight into dislocation dynamics can be gained by watching dislocation motion *in silico* in the three-dimensional crystal bulk when compared to a standard method of observing traces (steps) left by dislocations emerging at the two-dimensional crystal surfaces (as is often done in experiments). There are no physical surfaces available for slip trace observations in the full three-dimensional periodic boundary conditions in our MD simulations of an infinite crystal. To imitate surfaces, we chose to display atoms within certain "skin" distance from the initial nominal boundaries of the orthorhombic periodic box. At time t = 0, such skin atoms are arranged into nearly perfectly flat bounding surfaces. With the passage of time, steps appear on the fictional interfaces as dislocations pass through them and shift the skin atoms relative to each other. Such "slip traces" become increasingly more numerous and coarse under continuing tensile straining. We note that, unlike real (experimental) slip traces that are likely to reflect dislocation interactions with crystal surfaces, our imitation slip traces are signatures of dislocation activity in crystal bulk. Furthermore, whereas in experiments slip trace analyses are performed – with few exceptions – only in the end of straining tests, i.e. post mortem, “slip traces” extracted from our simulations are observed in silico in any degree of detail and at arbitrary time resolution. Still, looking inside the crystal bulk and observing every detail in the life of dislocations is much more informative.

## Acknowledgements
Authors acknowledge useful discussions with W. Cai, E. Tadmor and D. Karls and helpful editorial suggestions from D. Bulatova. This work was funded by the NNSA ASC Program and Technische Universität Darmstadt and was performed under the auspices of the US Department of Energy by Lawrence Livermore National Laboratory under contract W-7405-Eng-48. Computing support came from the DOE INCITE program and LLNL Computing Grand Challenge program. The simulations were performed on Mira and Vulcan supercomputers at the Argonne Laboratory Computational Facility and Livermore Computing Facility, respectively.

## Author contributions
L.A.Z.-R. and R.F. ran atomistic simulations, L.A.Z.-R. produced three atomistic movies, A.S. developed methods for *in silico* computational microscopy and visualization, T.O. optimized run-time efficiency and data management of ultra-scale simulations, N.B. developed algorithms for initialization of atomistic simulations, N.R.B. performed finite-element simulations and produced the “dogbone” movie, V.V.B. developed the concept, planned the research and generated starting configurations for molecular dynamics simulations. All authors analyzed simulation results and wrote the paper.

## Competing interests
The authors declare no competing interests.

## Data availability
All data presented in the main text and Supplementary Information is available at https://figshare.com. Large arrays of dislocation network data (around 400 GB in total) used for extracting dislocation densities and for producing supplementary movies are available from the corresponding author on a reasonable request.

## Code availability
The open source computer code LAMMPS used in this study is developed and maintained at the Sandia National Laboratories. LAMMPS is available at https://lammps.sandia.gov.

# Figure legends

**Fig. 1**: **Stress-strain response of an aluminum single crystal subjected to tensile straining along seven different initial orientations of the straining axis. a**, Stress-strain response extracted from MD simulations. Each line is labeled with the Miller index[11] of the crystal's initial axis orientation. The thin lines are raw stress-strain data, the thick lines are the same data smoothed using a moving average filter. The horizontal dashed lines are flow stress levels attained asymptotically as crystals approach their stable end orientations. **b**, Corresponding experimental stress-strain curves obtained in tensile straining tests of single crystal copper[3].

**Fig. 2**: **Slip crystallography of face-centered cubic single crystals. a**, **b**, **c**, Schematic illustration of crystal rotation due to slip on a single slip system under uniaxial tension. **d**, In the stereographic projection, point P representing an axis orientation on the unit sphere is projected onto point Q. A spherical triangle of axis orientations (bright green) is mapped onto a stereographic triangle on the equatorial plane (dark green). **e**, Slip direction (red arrow) and slip plane (gray) of one of the 12 slip systems in the face-centered cubic lattice. **f**, A 3D polar plot of the Schmid factor as a function of axis orientation with respect to the slip direction (red arrow) and the slip plane normal (blue arrow). **g**, The outer envelope of Schmid factors of all 12 slip systems in an FCC crystal under uniaxial straining. Plotted in 12 distinct colors, each of the 48 patches represents the amplitude of the Schmid factor on a slip system most favored for slip for a given axis orientation. The grooves coincide with the edges of the standard triangle where two slip systems are equally favored. At the intersection of the grooves lie three types of cusp orientations in which four, six or eight slip systems are equally favored, corresponding to high-symmetry axis orientations. **h**, Rotation trajectories of seven crystals shown in Fig. 1a. Placed next to the starting point of each trajectory are the Miller indices of the initial axis orientation.

F**ig. 3: Stress-strain response, dislocations densities and Schmid factors in 12 slip systems of aluminum. a**, Stress versus strain, **b**, dislocation densities versus strain and **c**, Schmid factors versus strain computed under tensile straining of a crystal with straining axis initially aligned with [101]. Thick lines are densities of the primary and the conjugate slip systems. All other slip systems are shown as thin lines. **d**, **e**, **f**, The same curves computed for a crystal strained along its [001] axis. Thick lines are densities of eight primary systems and four inactive slip systems are shown in thin lines. Line colors in b, c, e and f are coordinated with each other and with the colors of corresponding spherical triangles in Fig. 2g.

# Tables

**Table 1: Qualitative characteristics of strain hardening observed in MD simulations.**

| Initial axis | Initial slip symmetry | Crystal rotates? | Hardening response | End axis |
|---|---|---|---|---|
| [001] | 8-fold, holds* | no | parabolic | [001] |
| [111] | 6-fold, holds | no | parabolic | [111] |
| [101] | 4-fold, breaks* | yes | 3-stage | [112] |
| [112] | 2-fold, holds | no | parabolic | [112] |
| [212] | 2-fold, holds | yes | 3-stage | [111] |
| [102] | 2-fold (breaks | yes | 3-stage | [112] |
| [213] | no symmetry | yes | 3-stage | [112] |
| [8,5,13]** | no symmetry | yes | 3-stage | [112] |

* “Breaks” and “holds” indicate, respectively, whether or not the initial crystal orientation experiences symmetry-breaking under straining.

**Not shown in Fig.1A, stress-strain response of this crystal orientation is very similar to the [213] case.

## Supplementary information submitted in separate files

SI.pdf

Supplementary Discussions 1, 2, 3, 4, 5
Supplementary Figures 1, 2, 3, 4
Supplementary Tables 1, 2
Captions to Supplementary Movies 1, 2, 3, 4

Supplementary Movie 1
Supplementary Movie 2
Supplementary Movie 3
Supplementary Movie 4

# Supplementary Information

# Metal hardening in atomistic detail

Luis A. Zepeda-Ruiz, Alexander Stukowski, Tomas Oppelstrup, Nicolas Bertin, Nathan R. Barton, Rodrigo Freitas, Vasily V. Bulatov

**Supplementary Discussion 1**

Analysis of axis stability and rotation under uniaxial tension

One way to understand why symmetry breaks for some initial axes but does not break for the others is to examine *a priori* stability conditions for axis orientations on the edges and at the corners of the standard triangle. It is useful to recall that under uniaxial tension crystal axis rotates towards the most active slip direction(s), rotation rate being proportional to the instantaneous slip rate in that system. When multiple slip systems are simultaneously active, their rotation rate vectors add together. The three corner orientations do not rotate not because their active slip systems produce no slip, but because exactly equal amounts of slip are produced in each one of them and, by symmetry, their rotation rate vectors cancel each other.

Supplementary Figure 1 depicts *a priori* stability against symmetry-breaking and subsequent rotation for symmetric axis orientations on the edges and at the corners of the triangle. The arrows show which way the axes rotate when one of the equally favored slip systems slightly "outperforms" the other system(s) and produces a small amount of excess (uncompensated) slip. If an excess rotation causes reduction in the Schmid factor of the runaway system, the axis is predicted to be stable and unstable otherwise. According to this schematic, axes on the [001]—[101] and [101]—[111] edges should be unstable – these two triangle edges are "ridges". In contrast, all axes on the [001]—[111] edge should be stable against symmetry breaking thus defining a "groove". Anywhere on this groove, simultaneous and equal slip in two dominant systems – 21 (black) and 12 (red) - produces rotations that largely cancel each other, with a small remaining rotation component pointing towards the [112] orientation in the middle of the groove. Consequently, [112] axis orientation is predicted to be stable which is exactly what is observed both in experiments[3] and in our MD simulations. The [101] corner orientation falls on the intersection of two ridges and is unstable with respect to any one of its four active slip systems taking over, again in full agreement with our MD simulations (see main text) although experimental data on this orientation is scarce and inconclusive. Similar analysis predicts that axes [001] and [111] should be both stable under uniaxial tension, however their stability conditions are somewhat different. The [111] axis is "robustly stable" meaning that reduction in the Schmid factor of a runaway system caused by the rotation it causes is linearly proportional to the excess slip rate. By comparison, the [001] orientation is "delicately stable" meaning that the reduction in the Schmid factor of its runaway system is proportional only to the square of its excess slip rate. By the same measure, the [112] axis is delicately stable while the [101] axis is robustly unstable. All these qualitative conclusions are generally borne out by our MD simulations and experiment[34].

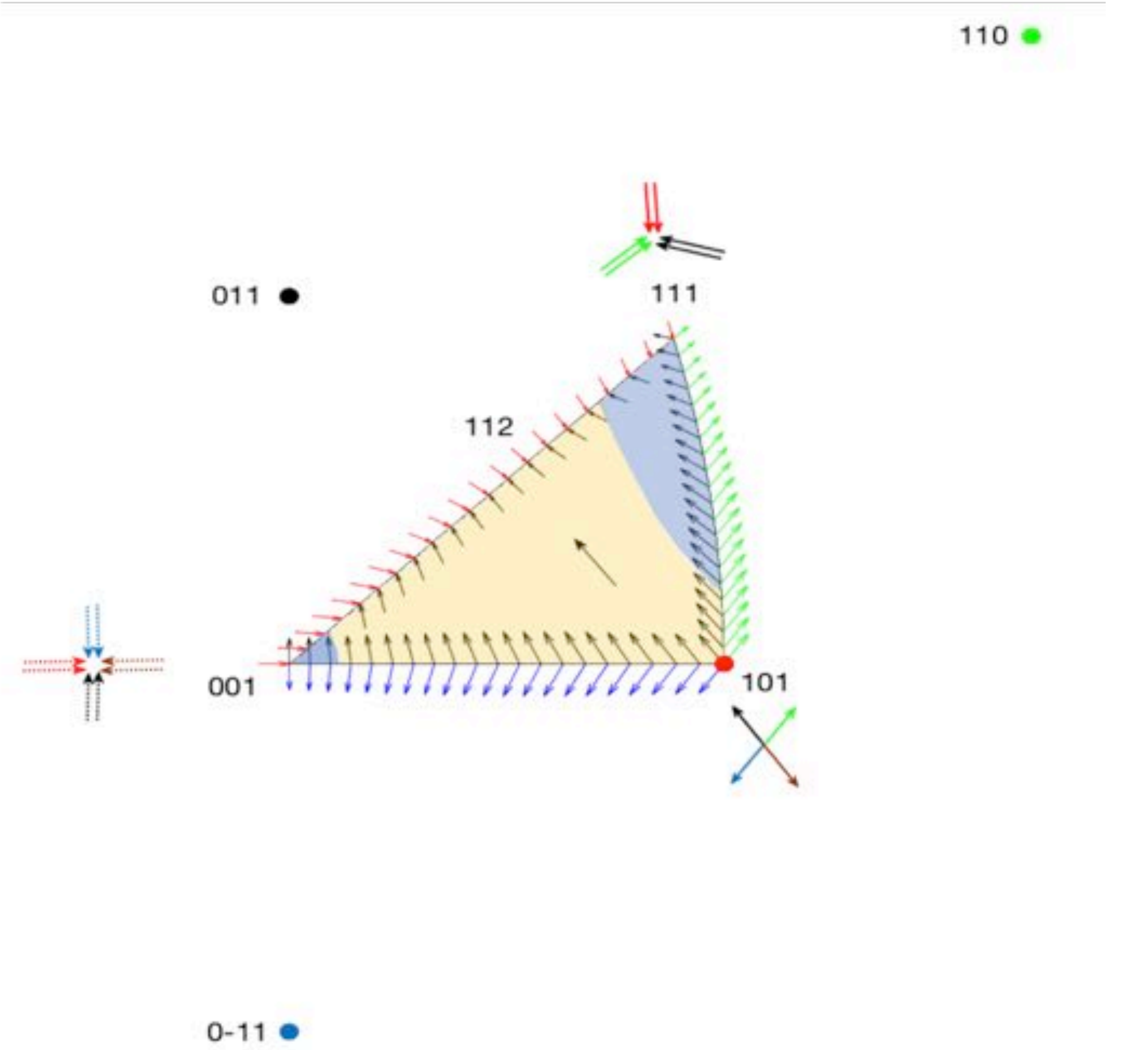

**Supplementary Figure 1:** Stereographic projections of rotation rates associated with slip in the most active slip systems, as a function of the tensile axis orientation. Each rotation vector points to the 001 pole projection of the slip direction (Burgers vector) causing the rotation, each pole plotted as a filled circle in the same color as the corresponding rotation vectors. The brown arrows point to slip directions that fall outside the area of stereographic projection shown in the figure. Length of each rotation vector is proportional to the rotation angle induced by a unit amount of slip in the corresponding slip system. The arrows show sense of rotation – towards or away from the triangle edges. The figures near the corners of the triangle illustrate stability of three corner orientations with respect to symmetry breaking, each double arrow depicting a pair of co-linear systems (with the same Burgers vector but different planes). The [001] and [111] axis orientations are stable and the [101] orientation is unstable. The dotted arrows pointing towards [001] indicate that this corner orientation is delicately stable. System 21 (black circle and arrows) is dominant everywhere inside the standard stereographic triangle drawn in the center. Three sub-areas of the standard triangle shaded in gray and yellow approximately delineate regions of attraction centered on three stable axis orientations 001, 112 and 111.

Qualitative predictions for axis stability and rotation depicted in Supplementary Figure 1 agree with all our MD simulations with one prominent exception: instead of rotating to the [112] orientation as suggested by this rotation flow schematic, [212] axis rotates towards and settles at [111] as its end orientation. Along the way, axis rotation trajectory never ventures far from the [101]—[111] edge predicted to be an unstable ridge (compare Supplementary Figure 1 and Fig. 2h in the main text). This finding is in agreement with a low-rate experiment reported in (*3*) where such rotation along the edge towards [111] was explained by a simultaneous and equal slip in two co-planar systems 21 (black) and 23 (green). In our view this explanation is not satisfactory: when two dominant systems are co-planar one would expect reduced forest resistance to dislocation motion (because dislocations moving in parallel planes do not have to cut through each other) and, thus, a relatively low hardening rate. Yet hardening rates observed in our MD simulation and in experiment[3] for the [212] initial axis orientation are high whereas co-planarity of systems 21 and 23 does not explain why this axis continues to rotate along the [101]—[111] edge despite its predicted instability. At present we do not have a convincing explanation for this peculiar axis rotation trajectory and its high hardening rate, however the very fact that our predictions are much in line with experiment suggest that our MD simulations capture the same physics of crystal plasticity. Having access to the entire MD trajectory and armed with our recently developed methods of *in silico* computational microscopy[28,29], we can track down origins of every wiggle in the stress-strain and density-strain curves to the underlying events in the life of atoms and dislocations, including the origin of the anomalous axis rotation from the [212] initial orientation. This we leave for future work.

In addition to eight axis orientations discussed in the main text, we performed our computational straining experiments on six other initial axis orientations. When strained to a sufficiently large strain (100 - 200 %), still only three initial axis orientations are found to be stable while the remaining 11 axes eventually rotate towards one of the three stable orientations, [001], [112] or [111]. This observation itself and our ability to reach very large strains in MD simulations are significant. Laboratory tests, such as shown in Fig. 1b, are inevitably interrupted by macroscopic instabilities, such as necking (under tension) and barreling (under compression) and the resulting stress-strain curves never extend to strains sufficient to observe asymptotic saturation. Nevertheless, it is important to understand how crystals deform and rotate at very large strains characteristic of various industrial methods of metal processing, e.g. rolling, channel die compression, drawing, extrusion, etc. In particular, asymptotic plasticity must surely affect texture development in polycrystals subjected to very large strains. Unlike experimental tests, our MD simulations are performed in three-dimensional periodic boundary conditions where instabilities such as necking (under tension) or barreling (under compression) are suppressed due to the absence of surfaces needed for instability initiation. The reason why some of our simulations did not fully reach flow saturation is mostly because large strains needed to convincingly observe such asymptotic behaviors demand very large simulations volumes (beyond our means) and large initial aspect ratios to avoid eventually thinning our atomistic models into a wire. Furthermore, in our simulations large numbers of vacancies and vacancy clusters are generated by the motion and intersections of dislocations. While almost all vacancies are observed to emerge in the form of the stacking fault tetrahedra[32], it is conceivable that further accumulation of such defects under extended tensile straining can result in an eventual material rupture even in the absence of any surfaces.

Asymptotic convergence to one of three terminal axes shows itself both in the stress-strain curves (compare Fig. 1a and Table 1 in the main text) and dislocation density-strain curves (Fig.

3 and Supplementary Figure 3). Based on a limited number of initial orientations we have tested so far, Supplementary Figure 1 sketches the boundaries of the cones of attraction around each asymptotically stable orientation. It is worth noting that our admittedly limited attempts to delineate the cone of attraction associated with the stable [001] orientation have not succeeded although we tested a few initial axis orientations rather close to [001], e.g. [104]. This is consistent with our qualitative analysis suggesting that [001] is a delicately stable axis. In contrast, the [111] axis is predicted to be robustly stable in agreement with our simulations showing that axes from a wide cone of initial orientations asymptotically rotate to [111]. Although uniaxial tension simulations reported here only sample one specific and limited portion of the full 5D space of deviatoric stress conditions, it appears plausible that terminal asymptotic behaviors may also be observed in more complex straining tests, e.g. bi-axial, multi-axial, tension versus compression, etc., in the limit of large strains.

**Supplementary Discussion 2**

Attribution of dislocations to slip systems

The DXA algorithm not only extracts dislocations from atomistic snapshots, but also indexes the same dislocations by assigning each of them a specific Burgers vector which must be equal to a translation vector of the crystal lattice (for complete dislocations) or to a certain fraction of a lattice translation vector (for fractional dislocations). In aluminum and other FCC crystals nearly all dislocations are fractional and are commonly attributed to one of two sub-categories: Shockley partials and stair rods. In the simulations discussed here we observe numerous stair-rods whose density steadily increases with strain. Nearly all stair rod dislocations have Burgers vectors of the 1/6<110> type and are nothing but edges of stacking fault tetrahedra (SFT) defects that result from dragging dislocation jogs and dislocation intersections[32]. Once formed, SFTs no longer move and participate in crystal plasticity only as obstacles to mobile Shockley partial dislocations. The latter dislocations have Burgers vectors of the 1/6<211> type and exist in the form of 12 symmetry-related but geometrically distinct variants. It is on this latter type of dislocations that one has to focus to understand plasticity response of aluminum and other FCC crystals.

Because the Burgers vector of a Shockley partial dislocation is only a fraction of a lattice translation vector, such partials always appear in pairs such that the Burgers vectors of two partials in every pair sum to a complete lattice vector of the ½<110> type, e.g.

$$\frac{1}{6}[11\bar{2}] + \frac{1}{6}[2\bar{1}\bar{1}] = \frac{1}{2}[10\bar{1}] \ .$$

Although two Shockley partials comprising a pair can and do move in their common glide plane somewhat independently of each other, the pair is glued together by a planar structural defect called stacking fault (SF). In most FCC metals, including aluminum, the SF is sufficiently wide to be observed in high-resolution electron microscopy experiments, yet still narrow enough for the motion of two partials to be closely coupled. Thus, it is the net force on the pair of Shockley partials that matters most in defining how a given pair will move.

The net Burgers vector of each given pair of Shockley partials is fully defined by the sum of their partial Burgers vectors, however the same complete Burgers vector of the ½<110> type can be a sum of two different pairs of partial Burgers vectors. For the same net Burges vector as in the previous example, there is an alternative pair of Shockley partials

$$\frac{1}{6}[21\bar{1}] + \frac{1}{6}[1\bar{1}\bar{2}] = \frac{1}{2}[10\bar{1}] \ .$$

This non-uniqueness is directly related to the existence of two alternative glide planes for each complete dislocation of the ½<110> type. The same non-uniqueness would make direct attribution of complete dislocations (as extracted by DXA) over 12 FCC systems difficult, because DXA yields only the Burgers vector but not the glide plane information of a dislocation. Fortunately, nearly all ½<110> dislocations come dissociated into pairs of Shockley partials in the glide planes and, thus, the slip system of any such Shockley pair is uniquely defined: for the first pair shown above the glide plane is (111), whereas for the second alternative pair the glide plane is $(1\bar{1}1)$. Thus, glide planes of complete ½<110> dislocations are inferred from the glide plane of their constituent partials. But here we face yet another non-uniqueness: while the glide plane of each Shockley pair is uniquely defined, viewed on its own each Shockley partial can be a ½-component of two different complete dislocations within the same glide plane. For example, partial $\frac{1}{6}[11\bar{2}]$ that appears in the very first pair listed above, can also be part of a different pair, i.e.

$$\frac{1}{6}[11\bar{2}] + \frac{1}{6}[\bar{1}2\bar{1}] = \frac{1}{2}[01\bar{1}] \ .$$

In FCC crystals, complete ½<110>dislocations dissociate and move in four glide planes of the {111} type, three complete dislocation per glide plane for a total of 12 glide-plane – Burgers vector combinations (slip systems). To streamline bookkeeping of the glide planes and to better explain how we apportion 1/6<112> partial dislocations over three different complete ½<110> dislocations in each of the four {111} planes, here we use a two-index numbering convention for complete and partial dislocations in an FCC lattice as an alternative to the widely accepted alphanumeric convention of Schmid and Boas[18] and the Latin-Greek alphabetic convention based on the Thompson tetrahedron (Supplementary Figure 2). Equally compact, our numbering convention is especially convenient for indexing the FCC slip systems as components of a vector or an array in a computer code and is easy to remember given its simple mnemonic rule for labeling the Burgers vectors and the glide planes. In our convention, the first index *k* (= 0, 1, 2, 4) in the complete and partial dislocation components – denoted $c_{kj}$ and $p_{kj}$ respectively - specifies one of the four glide planes of the {111} set: specifically, *k* corresponds to the component of the plane Miller index vector whose sign is different from the sign of two remaining Miller index components. For example, plane *k* = 2 has the sign of its second Miller index component different from the sign of its first and third components, which corresponds to plane $(1\bar{1}1)$ or, since the overall sign is immaterial, plane $(\bar{1}1\bar{1})$. By the same convention, $k = 0$ labels plane $(111)$ or $(\bar{1}\bar{1}\bar{1})$, The second index $j$ in $c_{kj}$ corresponds to the position of the "0" Miller index component of the Burgers vector of the ½<110> crystallographic family of complete dislocations. Likewise, the second index $j$ in $p_{kj}$ corresponds to the position of the "2" Miller index component in the 1/6<211> Burgers vector of the Shockley partial dislocations. For example, $c_{13}$ labels complete dislocations $\frac{1}{2}[110]$ in the $(\bar{1}11)$ plane whereas $p_{02}$ labels partial dislocations $\frac{1}{6}[1\bar{2}1]$ in the $(111)$ plane. As always, the Burgers vector of a dislocation must be perpendicular to the normal vector of its glide plane as specified by the plane's Miller index.

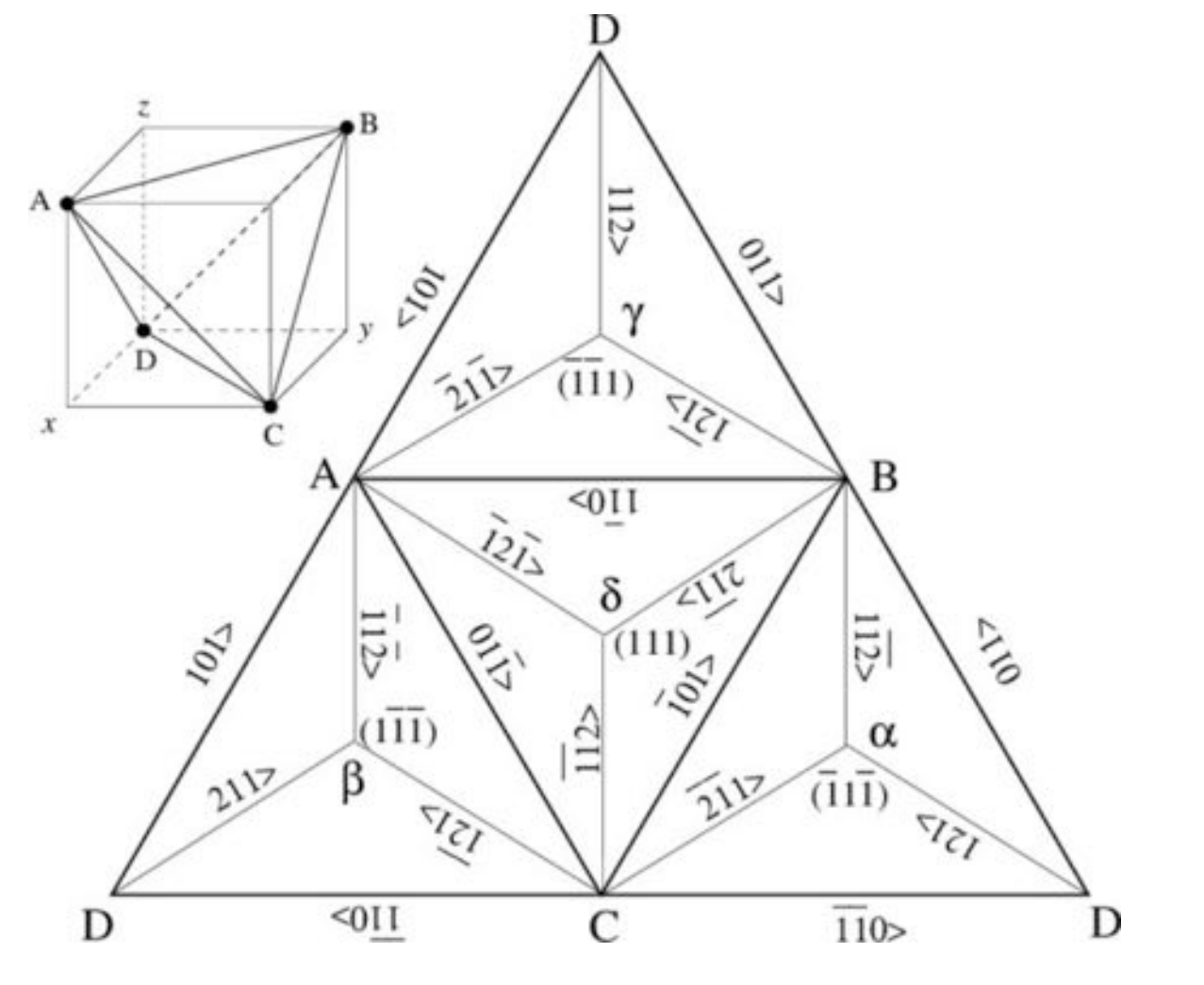

**Supplementary Figure 2:** Thompson tetrahedron notation for the Burgers vectors of complete and partial dislocations, slip planes and slip systems.

**Supplementary Table 1: Miller indices, Schmid and Boas labels, Thompson tetrahedron labels and our labeling convention for complete dislocations and slip systems in FCC crystals.**

| Miller index | Schmid and Boas label | Thompson tetrahedron label | Our label |
|---|---|---|---|
| $1/2[01\bar{1}](111)$ | A6 | AC | 01 |
| $1/2[10\bar{1}](111)$ | A3 | BC | 02 |
| $1/2[1\bar{1}0](111)$ | A2 | AB | 03 |
| $1/2[01\bar{1}](\bar{1}11)$ | B6 | AC | 11 |
| $1/2[101](\bar{1}11)$ | B4 | AD | 12 |
| $1/2[110](\bar{1}11)$ | B1 | CD | 13 |
| $1/2[011](1\bar{1}1)$ | C5 | BD | 21 |
| $1/2[10\bar{1}](1\bar{1}1)$ | C3 | BC | 22 |
| $1/2[110](1\bar{1}1)$ | C1 | CD | 23 |
| $1/2[011](11\bar{1})$ | D5 | BD | 31 |
| $1/2[101](11\bar{1})$ | D4 | AD | 32 |
| $1/2[1\bar{1}0](11\bar{1})$ | D2 | AB | 33 |

**Supplementary Table 2: Miller indices, Schmid and Boas plane labels, Thompson tetrahedron labels and our convention for 12 partial dislocations and slip systems in FCC crystals.**

| Miller Index | Schmid and Boas label (plane only) | Thompson tetrahedron label | Our label |
|---|---|---|---|
| $1/6[\bar{2}11](111)$ | A | Bδ | 01 |
| $1/6[1\bar{2}1](111)$ | A | Aδ | 02 |
| $1/6[11\bar{2}](111)$ | A | Cδ | 03 |
| $1/6[211](\bar{1}11)$ | B | Dβ | 11 |
| $1/6[12\bar{1}](\bar{1}11)$ | B | Cβ | 12 |
| $1/6[1\bar{1}2](\bar{1}11)$ | B | Aβ | 13 |
| $1/6[21\bar{1}](1\bar{1}1)$ | C | Cα | 21 |
| $1/6[121](1\bar{1}1)$ | C | Dα | 22 |
| $1/6[\bar{1}12](1\bar{1}1)$ | C | Bα | 23 |
| $1/6[2\bar{1}1](11\bar{1})$ | D | Aγ | 31 |
| $1/6[\bar{1}21](11\bar{1})$ | D | Bγ | 32 |
| $1/6[112](11\bar{1})$ | D | Dγ | 33 |

Supplementary Table 1 lists Miller indices of all twelve slips systems of complete dislocations in FCC crystals and their labels in the Schmid and Boas, Thompson tetrahedron and our own labeling conventions. Supplementary Table 2 lists Miller indices of all twelve partial slip systems in FCC, their Schmid and Boas plane labels, their alphabetic labels in the Thompson tetrahedron convention and their two-index labels in our labeling convention. We use our two-index notation only in the Supplementary Text to facilitate a more detailed discussion of slip system evolution during straining, whereas in the main text slip systems are referred to only by colors used to plot their densities and Schmid factors in Fig. 3 of the main text and Supplementary Figure 3.

Assuming that all complete dislocations come dissociated into Shockley pairs (which is fully borne out by our DXA data), the total lengths of three complete dislocations $c_{k1}$, $c_{k2}$, and $c_{k3}$ in each of the four glide planes $k$ (= 0, 1, 2, 3) is computed by solving 3 x 3 sets of linear equations

$$c_{k2} + c_{k3} = p_{k1}$$
$$c_{k1} + c_{k3} = p_{k2}$$
$$c_{k1} + c_{k2} = p_{k3}$$

where $p_{k1}$, $p_{k2}$, and $p_{k3}$ are total lengths of three partial dislocations in the same plane $k$ ( $p_{kj}$ are extracted by the DXA algorithm). The above equations neglect a small fraction of complete dislocations that do not dissociate into partials as well as a similarly small fraction of complex dislocation configurations such as Lomer-Cottrell locks[33] occasionally observed in our simulations.

Plotted in Fig. 3 of the main text, densities of compete dislocations in each of the 12 slip systems were obtained by solving four 3 x 3 linear sets of equations (one set for each glide plane) and dividing the solutions $c_{kj}$ for the line lengths of complete dislocations by the crystal volume.

**Supplementary Discussion 3**

Strain hardening as a function of initial axis orientation

Shown in Supplementary Figure 3 are simulated responses to tensile straining of aluminum single crystals of five initial axis orientations not included in Fig. 3 of the main text. The format of the plots is the same as in Fig. 3. Similar to the [101] crystal orientation shown in Fig. 3a, crystals with initial axis orientations [213], [102], and [212] show three-stage hardening or curve inflection (left panel of the upper row). Under straining these three crystals rotate along trajectories shown in Fig. 2h towards specific crystallographic end orientations listed in the fifth column of Table 1. One of these rotation trajectories is shown in Supplementary Movie S3 synchronized with the stress-strain curve, dislocation motion and plastic deformation of the simulation volume. In common to all three crystals, initial hardening rates (slopes of the stress-strain curve) are low ("easy glide"). Subsequent rise in the hardening rates is closely coordinated with the specimen axes rotating toward their end orientations. On approaching their end orientations – the [112] end orientation for the [213] and [102] crystals and the [111] end orientation for the [212] crystal – the flow stress appears to saturate. To understand precisely how crystal rotation and staged hardening are connected we now look into details of slip activity taking place under straining. Shown on the same figure as functions of strain just below the stress-strain curves are the geometric Schmid factors and dislocation densities in all 12 slip systems. To reduce clutter, the most heavily populated of the 12 slip systems are plotted in thick lines while the rest of them are shown in thin lines.

Straining response of cubic single crystals initially oriented along the [213] axis is often pointed to in the literature as exhibiting quintessential three-stage hardening. In this and any other axis orientation strictly inside (not on the edges or corners) of the standard triangle, one slip system - traditionally referred to as "primary" - is most favorably inclined for slip under the action of stress that develops in response to uniaxial straining. Represented by the thick solid black lines in the middle and the right columns, the primary slip system 21 becomes active from the very beginning and its fractional line length grows rapidly while other slip systems remain dormant. Slipping along this primary system, the crystal gradually rotates so that its straining axis moves towards the active (primary) slip direction [011]. As the axis approaches the [001]—[111] edge, the Schmid factor of the "conjugate" system 12 (thick red lines) becomes comparable and, when the axis crosses the triangle edge, equals that of system 21. This is very similar to the [101] crystal shown in Fig. 3 (a,b,c), except that the latter crystal enters easy glide only after breaking its initial 4-fold slip symmetry. However, stage I of easy glide in the [213] crystal is not as extended since this crystal takes less strain to rotate towards the [001]-[111] triangle edge and to raise the Schmid factor of and activate the conjugate slip system. Other than that, the inflection in the stress-strain curve occurs at approximately the same (rotated) axis orientations as in the [101] crystal. Again, similar to the [101] crystal, the axis rotation trajectory of the [213] crystal first overshoots but then returns to and settles at its end [112] orientation where both stress and fractional dislocation densities gradually approach their asymptotic levels. In agreement with experiment, we observed similar staged hardening in simulations of other initial crystal orientations in the interior of the standard triangle, e.g. the [8,5,13] orientation that is not shown in Fig. 3 or Supplementary Figure 3 but only listed in the bottom row of Table 1.

Shown in the second from top row in the same figure are results of another straining simulation in which similar staged hardening (with an inflection) is observed in a crystal initially oriented along the [102] axis. This as well as all other axes on the edges of the fundamental triangle initially have two maximally favored slip systems with equal Schmid factors. When strained along the [102] axis, two systems – 21 (black) and 01 (blue) - are initially equally active but the black system happens to produce a bit more slip causing the axis to rotate away from the [001]—[101] edge which further increases the Schmid factor of the runaway system (black) and decreases the Schmid factor of the competing system (blue). Now that the initial symmetry is broken, the axis begins to rotate in response to slip in the now dominant black system causing the blue system to increasingly lag behind and to eventually become inactive. The axis trajectory initially veers some towards the [001] corner, but then turns to the [001]—[111] edge and slowly inches towards its asymptotic [112] end orientation. As the axis is approaching the triangle edge, the conjugate slip system 12 (red) becomes increasingly active causing inflection hardening very similar to the [213] case. Just like in straining of the [101] axis discussed in the main text, here symmetry-breaking precedes axis rotation and associated inflection hardening. However, despite its predicted instability, another crystal initially oriented along [212] axis (the third row from top) retains it two-fold slip symmetry during straining while still rotating towards its [111] end orientation and showing distinct inflection hardening. This last example shows that neither initial lack of symmetry nor symmetry-breaking are prerequisites of staged hardening, but crystal rotation is.

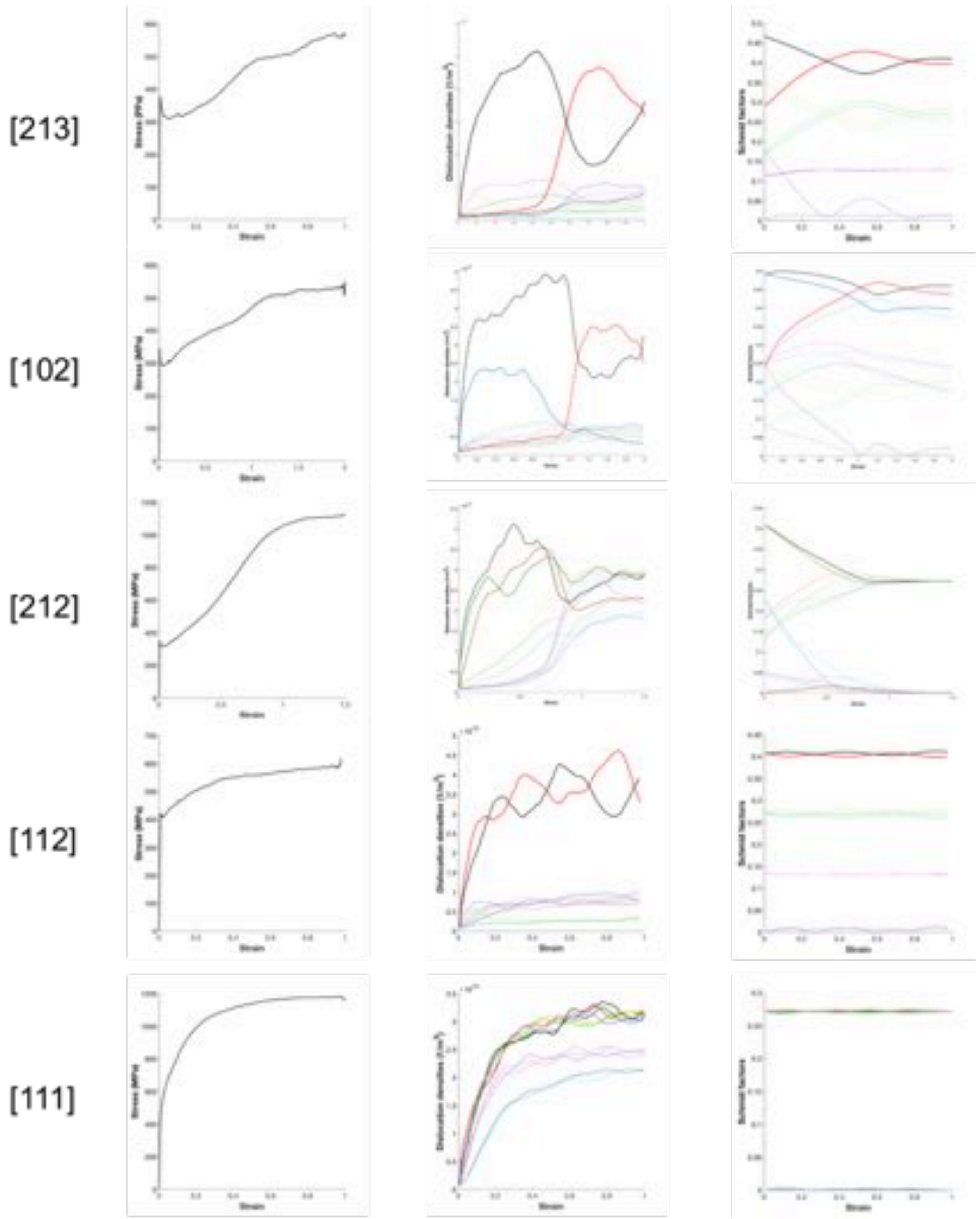

**Supplementary Figure 3:** Response to uniaxial straining of five crystals with different initial orientations of the straining axis. a, Stress-strain curves. b, Fractional densities of dislocations as functions of strain. c, Schmid factors as functions of strain. Miller indices of five initial crystal orientations are shown on the far left.

In addition to [001] whose straining response is shown in Fig. 3(D,E,F) of the main text, two other tested axis orientations – [112] and [111] - retain their initial symmetries, do not rotate and, as a result, show no staged hardening. Response to straining of the [112] crystal is simple (the fourth row from top): two initially favored systems 21 (black) and 12 (red) activate equally from the very beginning and remain the only two active systems along the entire straining trajectory. Small oscillations in the Schmid factors and more pronounced oscillations in the dislocation densities of two active systems are related to a low-amplitude (within two angular degrees) axis precession around the stable [112] orientation. The hardening response is parabolic. Response of the [111] crystal orientation (the bottom row) is similarly stable and parabolic except for a greater number (six) of active slip systems. Just as in the [001] case, we observe that dislocations in systems that have zero Schmid factors and are usually assumed to remain inactive, multiply nearly as much as in the active systems. Under [111] straining, fractional dislocation densities in six inactive systems are observed to fall into two groups: three systems reach higher densities while three other systems reach somewhat lower density levels (Fig. S3). The upper set of three inactive systems are co-planar with three pairs of active systems and may have been populated in the so-called co-planar and/or glissile junction reactions *(27)*. The lower three inactive systems all belong to the slip plane perpendicular to the straining axis and mechanisms of their multiplication remain to be established.

**Supplementary Discussion 4**

Stress-strain response in copper versus aluminum

MD and DDD simulations of crystal plasticity aim to model macroscopic plasticity response resulting from the collective motion of large ensembles of interacting dislocations. We have previously observed[25] that, for an MD or DDD model to be statistically representative, the model should contain no fewer than 10 thousand dislocation links (Here, a link is defined as a dislocation line connecting together two network nodes[35]). In the context of MD and DDD calculations, finite size effects are a major concern given that the computational resources needed to reach the necessary ensemble size of ~10,000 dislocation links are quite enormous.

Our plan for this study was to simulate straining response of the seven crystal orientations that were described by Takeuchi in his classic paper[3]. Even though Takeuchi's experiments were on copper for which a number of high-quality interatomic potentials exist, we opted to simulate aluminum largely predicated on our concern that in copper – given its low stacking fault energy (SFE) – dislocations would not multiply to the same extent as in aluminum thus possibly requiring larger simulation volumes than we could realistically afford even with the world class resources available to us. The higher SFE is known to increase the rate of dislocation cross-slip events thus promoting dislocation multiplication and increasing the number of links per unit volume[36-38]. Our concerns regarding copper were magnified once we learned of one prior attempt to simulate metal hardening in copper using MD (W. Cai, private communication). That attempt was unsuccessful ostensibly due to a shortage of dislocations in the simulated crystal volume likely resulting from one or both of the following: (1) a relatively small size of the simulated single crystal (~1 million atoms) which probably resulted in excessive self-annihilation of dislocations re-entering the volume through the periodic boundaries and (2) the low SFE of Cu known to impede dislocation cross-slip responsible for dislocation multiplication. For reference, SFE in aluminum is ~0.20 J/m$^2$ compared to only ~0.06 J/m$^2$ in copper.

To repeat Takeuchi's experiments *in silico* using atomistic MD simulations, we applied for and were awarded a large allocation of computational resources by the DOE INCITE program on ALCF Mira supercomputer. That initial award was for 110 million core hours which in the end was extended by additional 34% to complete our simulations. Yet even at the fully optimized performance of our production code LAMMPS, and with the vast amount of computer resources granted to us, each of our seven simulations had to be limited to < 300 million atoms, straining rates > $5 \cdot 10^7$/s and accumulated strains < 1.5. Given the constraints, we decided to simulate aluminum rather than copper to reduce probability of failure (to capture strain hardening). The resulting number of dislocation links attained in our MD simulations of aluminum ranged from 8 to 30 thousand (in the stage of a fully developed flow), sufficient for statistically relevant ensembles, but with not much of a margin.

In hindsight, were we justified in deciding to simulate aluminum instead of copper?

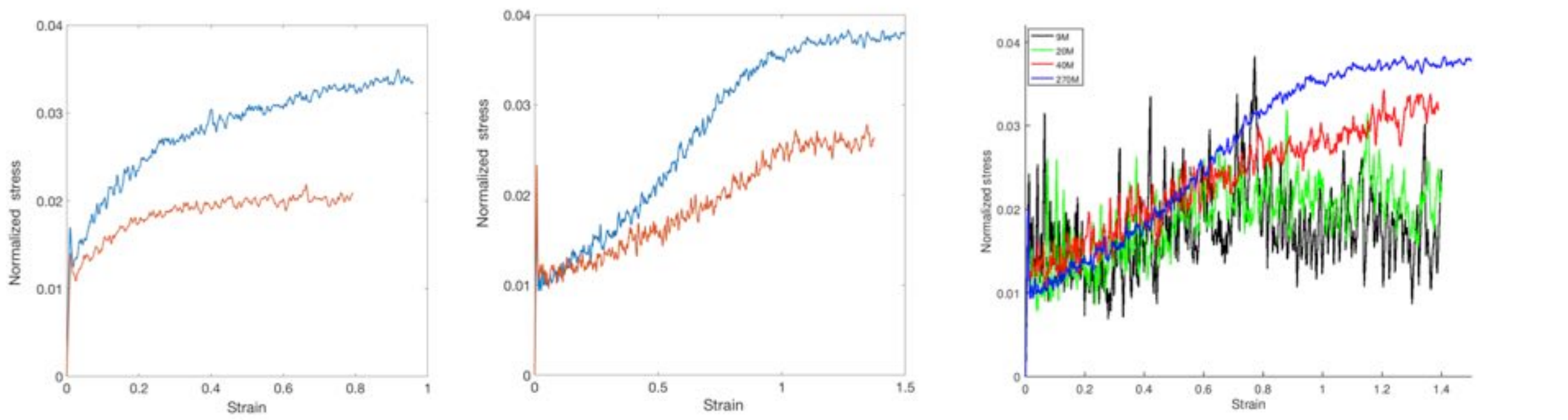

**Supplementary Figure 4:** Comparison of hardening response in copper ve rsus aluminum. a. Stress response to uniaxial tensile straining along the [001] axis in Cu (red) versus the same in Al (blue). Stress is normalized by the $G_{111}$ shear modulus for Cu and Al, respectively. b. Same as a, but now for straining along the [212] initial axis. c. Stress response of Al to straining along the [212] axis simulated in crystal volumes of different sizes: number of atoms for each curve are shown in the legend.

It has been amply documented in the literature that qualitative characteristics of metal hardening – the focus of this study – are very much the same in copper and aluminum as well as in other FCC metals[4-10]. Thus, to the extent that our simulations and experiments reflect the same physics and given our focus on qualitative behaviors, it should not have mattered which FCC metal to simulate. Nevertheless, to verify whether our initial concern about low SFE in copper was justified, 1.5 years after completing our seven simulations of aluminum, we performed a limited series of MD simulations on the Lassen supercomputer at LLNL under exactly the same simulation conditions as before, but now on copper. For the copper simulations we employed an interatomic potential that was previously shown to reproduce with high accuracy the properties of copper relevant to metal plasticity[39]. Gratifyingly, we observe qualitatively the same hardening behavior in copper as before in aluminum.

Shown in Supplementary Figure 4 are stress-strain curves for [001] and [212] crystal orientations selected here for comparison for their well-defined parabolic ([001)] and inflected ([212]) hardening curves observed in aluminum. The parabolic and the inflected shapes of the corresponding stress-strain curves in copper are unmistakable. Furthermore, the crystals are observed to rotate (the [212] crystal) or not to rotate (the [001] crystal) in exactly the same ways in copper as before in aluminum. At the same time, hardening in copper is not as well defined as in aluminum. To facilitate comparison, figure 4 shows flow stress divided (normalized) by an appropriate elastic shear modulus of copper and aluminum, respectively. For this normalization we used $G_{111}$, the shear modulus that defines stress response to shears in the {111} planes along the <110> slip directions[40]

$$G_{111} = (c_{11} - c_{12} + c_{44})/3$$

Here, $c_{11}$, $c_{12}$, and $c_{44}$ are the standard cubic elastic constants previously computed at T=0K for the interatomic potential models of aluminum and copper used in this study[24,43]. The resulting shear moduli are $G_{111}(Cu) = 41.2$GPa and $G_{111}(Al)$ = 29.7GPa.

After normalization by their respective $G_{111}$'s, the lower yield stresses attained in two metals right after the initial yield peaks happen to be very close to each other suggesting that our use of $G_{111}$ for normalization was indeed appropriate. However hardening in copper is markedly weaker than in aluminum, reaching normalized flow stress of 0.020 in Cu versus 0.034 in Al (and still hardening) for the [001] crystal and 0.026 in Cu versus 0.037 in Al for the [212] crystal. Additional analysis reveals that dislocation densities in copper are systematically lower than dislocation densities in aluminum across all stages of straining. For example, at strain 0.7 in the [001] simulation, total density of all ½<110> dislocations in copper is $1.8 \cdot 10^{16}/m^2$ versus $3.2 \cdot 10^{16}/m^2$ in aluminum. According to the celebrated Taylor law[13], square root of the ratio of dislocation densities in Al and Cu should be approximately equal to the square root of the ratio of their normalized flow stresses, all taken at the same stage of straining. Indeed, the two ratios are not far apart, at 1.34 for the densities and ~1.5 for the stresses. Thus, it appears that dislocation under-multiplication in copper is largely responsible for the lower hardening.

Also pointing to relative scarcity of dislocations in copper are the greater amplitude of stress fluctuations and the higher peak yield stress in copper – both differences stand out more clearly when looking at the raw stress-strain data (before normalization) not shown here. Following the same trend as dislocation densities, the number of links attained in our copper simulations is correspondingly lower than in aluminum, exceeding our rule of thumb minimum of 10 thousand links only in one attempted simulation. To the extent that dislocation multiplication is defined by the ease of cross-slip, the lower SFE of copper must be responsible for the observed differences. To observe if and how straining response is affected by the size of the simulated crystal, we repeated some of our aluminum simulations in smaller periodic volumes. As an example, Supplementary Figure 3c shows simulated stress response of four model crystals subjected to tensile straining along the same [212] axis and containing 9, 20, 40 and 270 million atoms, respectively. With the decreasing size, hardening diminishes while the amplitude of stress fluctuations increases to the extent of nearly completely masking hardening response of the model with 8 million atoms.

Taken together, these observations bring us to conclude that taking copper as a test bed metal for our study would have worked too, but our concern about its low SFE was partially justified.

**Supplementary Discussion 5**

Stress state in a single crystal under uniaxial tension

Finite element calculations using a standard continuum crystal-mechanics-based constitutive model were used to assess the degree to which tension experiments produce uniaxial states of stress in the middle of the gauge section. These 3D calculations made use of a crystal plasticity model similar to those found in the literature[41-43]. The calculations confirm both a uniaxial state of stress at the middle of the gauge section and a tendency for deformation to localize into a "neck" region of the gauge section as the hardening rate falls off. This necking is consistent with expectations from standard metal formability considerations[44], and formation of a neck limits the amount of strain accumulated in a uniaxial stress condition. These finite element simulations also provided a means of confirming basic behaviors of the crystal lattice rotation under uniaxial stress deformation conditions. A movie of one of these simulations is provided (Supplementary Movie 4), showing both the stress response averaged over the middle of the gauge section and a graphical depiction of the alignment of the stress with the tensile direction. While the average stress in the middle of the specimen's gauge section remains uniaxial, one can see from the movie that significant stress heterogeneity develops in the neck region and in the dog bone shoulder sections.

**Captions to supplementary movies**

Supplementary Movie 1: Simulated response of single crystal aluminum to tensile straining along $x$ axis initially oriented along [001] direction of the cubic lattice. Two animated sequences at the top depict continuous elongation of the simulated crystal along the $x$ axis and simultaneous reduction of its lateral dimensions due to the Poisson effect. The sequence at the top left shows emergence and coarsening of “slip traces” on fictitious surfaces of the simulated crystal. The sequence at the top right shows dislocation motion and multiplication resulting in the development of a dense dislocation network. Sown at the bottom left is the stress-strain response of this crystal under tensile straining which is distinctly parabolic (no inflection). At the bottom right is the orientation trajectory of the tensile axis expressed in a frame tied to the cubic lattice of the strained crystal. The axis remains within 2 angular degrees from its initial [001] orientation during straining. All four animated sequences are synchronized. Axis orientations of the Laboratory (specimen) frame are shown at the far left.

Supplementary Movie 2: Simulated response of single crystal aluminum to tensile straining along $x$ axis initially oriented along [101] direction of the cubic lattice. Meaning of the four synchronized sequences is as explained in the caption to Supplementary Movie 1. In this case an initially orthorhombic crystal not only elongates along its straining axis $x$, but also changes its shape and becomes distinctly triclinic. Although not immediately obvious from changing orientations of the slip traces in the upper left sequence, axis rotation from its initial [101] orientation at the corner of the standard triangle towards [112] as well as axis overshoot, are both clearly shown at the bottom left. The stress-strain response shown at the bottom left is a typical staged (inflection) hardening.

Supplementary Movie 3: Simulated response of single crystal aluminum to tensile straining along $x$ axis initially oriented along [213] direction of the cubic lattice. Meaning of the four synchronized sequences is as explained in the caption to Supplementary Movie 1. As shown at the bottom right, the axis rotates along the triangle edge towards [112] orientation. The stress-strain response shown at the bottom left shows a distinct inflection.

Supplementary Movie 4: Deformation of a “dogbone” specimen subjected to uniaxial straining in tension. (Left) Evolution of the stress tensor in the dogbone under tensile straining. Colors other than white indicate a misalignment between the first principal axis of the stress tensor (corresponding to its maximum eigenvalue) and the tensile axis. (Right) Six components of the stress tensor averaged over the middle portion of the specimen. Position of the vertical bar along the strain axis is synchronized with specimen deformation shown on left. All stress components but $\sigma_{xx}$ remain close to zero along the entire straining path.